\documentclass{webofc}
\usepackage[varg]{txfonts}   % Web of Conferences font
\newcommand{\be}{\begin{equation}}
\newcommand{\ee}{\end{equation}}
\newcommand{\bea}{\begin{eqnarray}}
\newcommand{\eea}{\end{eqnarray}}

\begin{document}
\title{Flattening the inflaton potential beyond minimal gravity}

\author{\firstname{Hyun Min} \lastname{Lee}\inst{1}\fnsep\thanks{\email{ hminlee@cau.ac.kr}}
        % etc.
}

\institute{Department of physics, Chung-Ang University, Seoul 06974, Korea. \\
          }

\abstract{%
  We review the status of the Starobinsky-like models for inflation beyond minimal gravity and discuss the unitarity problem due to the presence of a large non-minimal gravity coupling. We show that the induced gravity models allow for a self-consistent description of inflation and discuss the implications of the inflaton couplings to the Higgs field in the Standard Model. 
  
  \vspace{0.8cm}
  
\begin{centering}
{\it Prepared for the proceedings of the 13th International Conference on Gravitation, \\
\hspace{2.5cm} Ewha Womans University, Korea, 3-7 July 2017.}
\end{centering}

}

\maketitle
\section{Introduction}

It was precisely five years ago since the discovery of Higgs boson on July 4, 2012 \cite{Higgs} that I gave this talk in the 2017 International Conference on Gravitation. 
Together with the detection of gravitational waves coming from the merger of binary black holes on September 14, 2015 \cite{GW}, we have the fortune to encounter the memorable moments of triumphs of the Standard Model (SM) and General Relativity in our century. 
Even with such a huge success of our understanding fundamental interactions from subatomic scales to cosmological distances, there remain open questions on the validity of our theories at high energies and in the early Universe. 

Cosmological inflation  solves
problems of horizon, homogeneity, flatness, cosmic relics,
and structure formation, requiring the period of an exponential expansion with a scalar field ``inflaton'' just after Big Bang.  From the measurement of Cosmic Microwave Background (CMB) anisotropies by Planck satellite \cite{planck}, almost scale-invariant and Gaussian scalar perturbations are needed, meaning that  canonical single field inflation models are favored. Inflation could also lead to primordial gravity waves in the CMB polarizations at the detectable level in the future CMB experiments.  

Almost scale-invariant CMB anisotropies require the inflation potential to be flat for a long period of time, i.e. $N=60$ e-foldings. In most of large field inflation models, the inflaton field makes a trans-Planckian excursion during inflation.  
The question is then whether inflation potential remains flat for entire range of inflaton fields such that UV physics can be ignored.

In this article, we review on a class of inflation models beyond minimal gravity and discuss the unitarity problems due to a large non-minimal coupling. We show that in induced gravity models, the inflaton potential is justified against the quantum corrections due to physics below the Planck scale.

\section{Inflation beyond minimal gravity}

In Einstein gravity, the Lagrangian for a real scalar inflaton  $\phi$ is described by
\be
{\cal L}_{\rm E}=\sqrt{-g} \left(\frac{1}{2}{\cal R}-\frac{1}{2} (\partial\phi)^2 - V(\phi) \right).
\ee
In order to realize a sufficiently long period of inflation, we need to introduce small masses or couplings in the inflaton potential. For instance, we meed $m\sim 10^{13}\,{\rm GeV}$ for quadratic potential, $V(\phi)=\frac{1}{2}m^2 \phi^2$; $\lambda\sim 10^{-12}$ for quartic potential, $V(\phi)=\frac{1}{4}\lambda \phi^4$. 

In the case of scalar-tensor gravity, the general Lagrangian for a real scalar inflaton contains a non-minimal coupling $F(\phi)$ as below,
\be
{\cal L}_{\rm J}=\sqrt{-g} \left(\frac{1}{2}{\cal R}+\frac{1}{2}F(\phi){\cal R}-\frac{1}{2} (\partial\phi)^2 - U(\phi) \right).
\ee
Then, if the effective Planck mass is field-dependent as $M^2_P(\phi)\sim F(\phi)$ with $F^2(\phi)\sim U(\phi)$  during inflation, 
gravity becomes weaker and makes the effective inflation potential $V\sim U/F^2$
asympotically constant,  without a need of small couplings. 

It is known that higher curvature terms can be introduced to drive inflation without a scalar field, because the degree of freedom in gravity becomes then larger than two. 
The Starobinsky model \cite{Star} introduces an ${\cal R}^2$ term in the Lagrangian as below,
\be
{\cal L}_{\rm Star}= \sqrt{-g} \left(\frac{1}{2}{\cal R} +\frac{1}{16}\xi^2 {\cal R}^2 \right). 
\ee
Then, it can be shown that the model is dual to a scalar-tensor gravity at the classical level, after the introduction of a Lagrange multiplier $\phi$, as
\be
{\cal L}_{\rm dual}= \sqrt{-g} \left(\frac{1}{2}{\cal R} +\frac{1}{2}\xi f(\phi) {\cal R} - U(\phi) \right) 
\ee
for $f(\phi), U(\phi)$ satisfying $U(\phi)=f^2(\phi)$, which is called the Starobinsky condition. 
Thus, there are a class of scalar-tensor theories, that are equivalent to the Starobinsky model.  
As discussed previously, the dual theory automatically satisfies the condition for the asymptotic flat effective potential of the inflaton, due to the functional relation, $U(\phi)=f^2(\phi)$.
For instance, for $U(\phi)=\frac{1}{4}\lambda \phi^4$, we would need $f(\phi)\sim \phi^2$. 

As a result, after rescaling the metric and the scalar field for canonical kinetic terms by $g_{\mu\nu}\rightarrow g_{\mu\nu}/ \Omega$ with $\Omega=1+\xi f(\phi)$ and $\frac{d\chi}{d\phi}=\sqrt{\frac{3}{2}} \frac{\Omega'}{\Omega}$, respectively, the inflaton Lagrangian in the scalar dual theory becomes the following form,
\be
{\cal L}_{\rm dual}= \sqrt{-g_E} \left(\frac{1}{2}{\cal R}-\frac{1}{2}(\partial\chi)^2 -\frac{1}{\xi^2} \Big(1-e^{-2\chi/\sqrt{6}} \Big)^2 \right).  \label{starobinsky}
\ee
The Starobinsky model leads to inflationary predictions for the spectral index and tensor-to-scalar ratio, 
$n_s=1-\frac{2}{N}, r=\frac{12}{N^2}$, with $\xi\sim 10^4$ for the normalization of the CMB anisotropies.
Therefore, the results are well consistent with the Planck data \cite{planck}.
On the other hand, in the SM Higgs inflation with non-minimal coupling \cite{Higgsinf}, the Higgs kinetic term can be ignored during inflation, so it is classically equivalent to the Starobinsky model, leading to the similar predictions for inflation.  
However, the consistency of the Starobinsky model and the Higgs inflation below the Planck scale might be questioned at the quantum level \cite{Higgs-unitarity,Higgs-unitarity2}, due to the presence of a large non-minimal coupling $\xi$.

\section{Induced gravity and unitarity problem}

We consider a general gravity Lagrangian including the general inflaton kinetic term in the following \cite{gian},
\be
{\cal L}_{\rm gen}=\sqrt{-g} \left(\frac{1}{2}\Omega(\phi){\cal R}-\frac{1}{2}K(\phi)(\partial\phi)^2-U(\phi) \right).
\ee
Then, after rescaling the metric by $g_{\mu\nu}\rightarrow g_{\mu\nu}/ \Omega$, the above Lagrangian becomes in Einstein frame as
\be
{\cal L}_{\rm gen}=\sqrt{-g_E} \left(\frac{1}{2}{\cal R}-\frac{1}{2}\left( \frac{K}{\Omega}+\frac{3\Omega^{\prime 2}}{\Omega^2}\right)(\partial\phi)^2-V  \right)  \label{genL}
\ee
with  $V=U/\Omega^2$. 
Identifying the canonical inflation field by $\frac{d\chi}{d\phi}=\sqrt{K/\Omega+3\Omega^{\prime 2}/2\Omega^2}$ , the slow-roll condition is ensured for $\epsilon=\frac{1}{2}(dV/\chi /V)=\frac{1}{3}(\Omega U'/(\Omega' U)-2)^2\ll 1$, constraining $U(\phi)\sim \Omega^2 [1+{\cal O}(\sqrt{\epsilon})] $.  Furthermore, for Starobinky-like inflation, we take 
\be
\Omega^{\prime 2}/\Omega^2\gg K \quad \quad {\rm and} \quad\quad  \frac{1}{2}\Omega' R = U'.
\ee
As a result, we get $\Omega=1+\frac{R}{4V_I}$ and $U=\frac{R^2}{16V_I}$, fixing the potential completely \cite{gian} to
\be
U=V_I (\Omega-1)^2  \label{star-pot}
\ee
with $V_I$ being the vacuum energy during inflation. 
Then, during inflation, the Lagrangian (\ref{genL}) with eq.~(\ref{star-pot}) becomes the same form  as eq.~(\ref{starobinsky}) in the Starobinsky model, but it takes a more general form due to the general kinetic terms for inflation, thus leading to Starobinsky-like models.    

Choosing $K=1$ and the frame function $\Omega$ and the potential \cite{gian} as follows,
\be
\Omega= \xi f(\phi), \quad  U=\lambda\, \Big( f(\phi)-\xi^{-1} \Big)^2,  \label{induced}
\ee
we can generate the Planck mass by the inflaton VEV such that $f(\langle\phi\rangle)=\xi^{-1}$, as in the induced gravity model \cite{induced}. 
Then, for $f(\phi)=\phi^n$, the inflaton kinetic term in Einstein frame becomes in vacuum \cite{gian}
\be
\frac{{\cal L}_{\rm kin}}{\sqrt{-g_E}}= -\frac{1}{2} \left( \frac{K}{\Omega}+\frac{3\Omega^{\prime 2}}{2\Omega^2}\right)\bigg|_{\phi=\langle\phi\rangle}(\partial\phi)^2= -\frac{3}{4} n^2 \xi^{2/n} (\partial\phi)^2.
\ee
In this case, defining the canonical inflaton field $\chi$ in vacuum by $\phi=\sqrt{\frac{2}{3}} \frac{1}{n}\, \xi^{-1/n}\chi$, we obtain the inflaton potential in Einstein frame,  
$V=U/\Omega^2$, with $f(\phi)=\phi^n$, as
\be
V=\frac{\lambda}{\xi^2} \Big[1-(\xi \phi^n)^{-1}\Big]^2= \frac{\lambda}{\xi^2} \left[ 1-\left(\frac{3}{2} \right)^{n/2} n^n\, \chi^{-n} \right]^2. 
\ee
Therefore, after expanding the inflaton around $\langle\phi\rangle=\xi^{-1/n}$, we find that the interaction terms for the canonical inflaton field $\chi$ are suppressed by the Planck scale, so there is no violation of unitarity below the Planck scale \cite{gian}.  The induced gravity model with $f(\phi)=\phi^2$ and a mixing quartic coupling between $\phi$ and Higgs field has been proposed as a unitarization of the original Higgs inflation with non-minimal coupling \cite{unitarity}. According to the above discussion, we note that there are a general class of induced gravity models unitarizing the Higgs inflation. 
The discussion with a large inflaton VEV has been extended to the case with a general form of the kinetic term $K(\phi)$ \cite{kinterm}. 

On the other hand, in universal attractor models \cite{linde} where $K=1$, $\Omega=1+\xi f(\phi)$ and $U=\lambda f^2(\phi)$ are taken, a small or vanishing inflation VEV is assumed. In this case, although the inflationary predictions are the same as in induced gravity models, the non-minimal coupling does not rescale the inflaton field in vacuum such that the interaction terms in the inflaton potential are suppressed by $\Lambda_{\rm UV}=M_P/\xi^{1/(n-1)}$, which hints at the premature violation of unitarity below the Planck scale \cite{gian}.  This class of models contains the Higgs inflation with non-minimal coupling for $n=2$ \cite{Higgs-unitarity,Higgs-unitarity2}. We note that there is an exception to the low-scale violation of unitarity for $n=1$ \cite{riotto,espinosa}, in which case the inflaton VEV is not physical because it can be always absorbed by the shift of the inflation field. 

Finally, we remark the implications of the inflaton coupling in induced gravity model for vacuum stability and reheating. 
In the presence of both the inflaton $\phi$ and the SM Higgs scalar $h$, we can choose the non-minimal couplings and the scalar potential in eq.~(\ref{induced}) \cite{unitarity} as follows,
\bea
\xi f(\phi,h) &=& \xi_\phi \phi^2 + \xi_h h^2, \\
U(\phi,h) &=& \frac{1}{4}\lambda_\phi (\phi^2-w^2)^2 +\frac{1}{4} \lambda_h (h^2-v^2)^2 +\frac{1}{2}\lambda_{H\phi} (h^2-v^2)(\phi^2-w^2),
\eea
with the inflation VEV given by $w\sim1/\sqrt{\xi_\phi}\gg  v$. 
Then, it has been shown that a nonzero mixing quartic coupling $\lambda_{H\phi}$ improves the stability of electroweak vacuum by inducing a tree-level shift in the effective quartic coupling, $\lambda_{\rm eff}=\lambda_h-\frac{\lambda^2_{H\phi}}{\lambda_\phi}$ \cite{VSB1,VSB2}, which is inferred from the measured Higgs mass, $m_h=\sqrt{2\lambda_{\rm eff}}\, v$. 
Furthermore, the inflaton decay into a pair of Higgs bosons reheats the Universe after inflation, thanks to the mixing quartic coupling $\lambda_{H\phi}$, leading to the reheating temperature in the range of $10^9\,{\rm GeV}\lesssim T_R\lesssim 10^{13}\,{\rm GeV}$ \cite{reheat}.

\section{Conclusions}

We have shown that the inflaton potential can be made flat without a small coupling, as gravity becomes weaker due to an inflaton-dependent effective gravity coupling during inflation. 
We have revisited the Starobinsky model and its scalar-dual theories as well as the Higgs inflation with a large non-minimal coupling, all of which are consistent with Planck data.
Generalizing to Starobinsky-like models in scalar-tensor theories, we have identified the induced gravity models to be consistent at  the quantum level below the Planck scale.
We have also discussed the implications of the coupling between the inflaton and the Higgs field for vacuum stability and reheating.

%\begin{figure}[h]
%\centering
%\includegraphics[width=1cm,clip]{tiger}
%\label{fig-1}       % Give a unique label
%\end{figure}

\section*{Acknowledgments}

The author would like to thank Cliff Burgess, Jose Espinosa, Gian Giudice and Mike Trott for collaboration and discussion. 
The work is supported in part by Basic Science Research Program through the National Research Foundation of Korea (NRF) funded by the Ministry of Education, Science and Technology (NRF-2016R1A2B4008759).


\begin{thebibliography}{}


\bibitem{Higgs}
%\cite{Aad:2012tfa}
%\bibitem{Aad:2012tfa}
  G.~Aad {\it et al.} [ATLAS Collaboration],
  %``Observation of a new particle in the search for the Standard Model Higgs boson with the ATLAS detector at the LHC,''
  Phys.\ Lett.\ B {\bf 716} (2012) 1
  doi:10.1016/j.physletb.2012.08.020
  [arXiv:1207.7214 [hep-ex]];
  %%CITATION = doi:10.1016/j.physletb.2012.08.020;%%
  %7536 citations counted in INSPIRE as of 19 Aug 2017
%\cite{Chatrchyan:2012xdj}
%\bibitem{Chatrchyan:2012xdj}
  S.~Chatrchyan {\it et al.} [CMS Collaboration],
  %``Observation of a new boson at a mass of 125 GeV with the CMS experiment at the LHC,''
  Phys.\ Lett.\ B {\bf 716} (2012) 30
  doi:10.1016/j.physletb.2012.08.021
  [arXiv:1207.7235 [hep-ex]].
  %%CITATION = doi:10.1016/j.physletb.2012.08.021;%%
  %7367 citations counted in INSPIRE as of 19 Aug 2017



\bibitem{GW}
%\cite{Abbott:2016blz}
%\bibitem{Abbott:2016blz}
  B.~P.~Abbott {\it et al.} [LIGO Scientific and Virgo Collaborations],
  %``Observation of Gravitational Waves from a Binary Black Hole Merger,''
  Phys.\ Rev.\ Lett.\  {\bf 116} (2016) no.6,  061102
  doi:10.1103/PhysRevLett.116.061102
  [arXiv:1602.03837 [gr-qc]].
  %%CITATION = doi:10.1103/PhysRevLett.116.061102;%%
  %1675 citations counted in INSPIRE as of 19 Aug 2017



\bibitem{planck}
%\cite{Ade:2015lrj}
%\bibitem{Ade:2015lrj}
  P.~A.~R.~Ade {\it et al.} [Planck Collaboration],
  %``Planck 2015 results. XX. Constraints on inflation,''
  Astron.\ Astrophys.\  {\bf 594} (2016) A20
  doi:10.1051/0004-6361/201525898
  [arXiv:1502.02114 [astro-ph.CO]].
  %%CITATION = doi:10.1051/0004-6361/201525898;%%
  %1236 citations counted in INSPIRE as of 19 Aug 2017


\bibitem{Star}
%\cite{Starobinsky:1980te}
%\bibitem{Starobinsky:1980te}
  A.~A.~Starobinsky,
  %``A New Type of Isotropic Cosmological Models Without Singularity,''
  Phys.\ Lett.\  {\bf 91B} (1980) 99.
  doi:10.1016/0370-2693(80)90670-X
  %%CITATION = doi:10.1016/0370-2693(80)90670-X;%%
  %3263 citations counted in INSPIRE as of 19 Aug 2017


\bibitem{Higgsinf}
%\cite{Bezrukov:2007ep}
%\bibitem{Bezrukov:2007ep}
  F.~L.~Bezrukov and M.~Shaposhnikov,
  %``The Standard Model Higgs boson as the inflaton,''
  Phys.\ Lett.\ B {\bf 659} (2008) 703
  doi:10.1016/j.physletb.2007.11.072
  [arXiv:0710.3755 [hep-th]].
  %%CITATION = doi:10.1016/j.physletb.2007.11.072;%%
  %963 citations counted in INSPIRE as of 19 Aug 2017




\bibitem{Higgs-unitarity}
%\cite{Burgess:2009ea}
%\bibitem{Burgess:2009ea}
  C.~P.~Burgess, H.~M.~Lee and M.~Trott,
  %``Power-counting and the Validity of the Classical Approximation During Inflation,''
  JHEP {\bf 0909} (2009) 103
  doi:10.1088/1126-6708/2009/09/103
  [arXiv:0902.4465 [hep-ph]];
  %%CITATION = doi:10.1088/1126-6708/2009/09/103;%%
  %206 citations counted in INSPIRE as of 19 Aug 2017
%\cite{Burgess:2010zq}
%\bibitem{Burgess:2010zq}
  C.~P.~Burgess, H.~M.~Lee and M.~Trott,
  %``Comment on Higgs Inflation and Naturalness,''
  JHEP {\bf 1007} (2010) 007
  doi:10.1007/JHEP07(2010)007
  [arXiv:1002.2730 [hep-ph]].
  %%CITATION = doi:10.1007/JHEP07(2010)007;%%
  %176 citations counted in INSPIRE as of 19 Aug 2017


\bibitem{Higgs-unitarity2}
%\cite{Barbon:2009ya}
%\bibitem{Barbon:2009ya}
  J.~L.~F.~Barbon and J.~R.~Espinosa,
  %``On the Naturalness of Higgs Inflation,''
  Phys.\ Rev.\ D {\bf 79} (2009) 081302
  doi:10.1103/PhysRevD.79.081302
  [arXiv:0903.0355 [hep-ph]].
  %%CITATION = doi:10.1103/PhysRevD.79.081302;%%
  %232 citations counted in INSPIRE as of 19 Aug 2017
%\cite{Hertzberg:2010dc}
%\bibitem{Hertzberg:2010dc}
  M.~P.~Hertzberg,
  %``On Inflation with Non-minimal Coupling,''
  JHEP {\bf 1011} (2010) 023
  doi:10.1007/JHEP11(2010)023
  [arXiv:1002.2995 [hep-ph]].
  %%CITATION = doi:10.1007/JHEP11(2010)023;%%
  %155 citations counted in INSPIRE as of 19 Aug 2017




\bibitem{gian}
%\cite{Giudice:2014toa}
%\bibitem{Giudice:2014toa}
  G.~F.~Giudice and H.~M.~Lee,
  %``Starobinsky-like inflation from induced gravity,''
  Phys.\ Lett.\ B {\bf 733} (2014) 58
  doi:10.1016/j.physletb.2014.04.020
  [arXiv:1402.2129 [hep-ph]].
  %%CITATION = doi:10.1016/j.physletb.2014.04.020;%%
  %38 citations counted in INSPIRE as of 19 Aug 2017


\bibitem{induced}
%\cite{Zee:1978wi}
%\bibitem{Zee:1978wi}
  A.~Zee,
  %``A Broken Symmetric Theory of Gravity,''
  Phys.\ Rev.\ Lett.\  {\bf 42} (1979) 417;
  doi:10.1103/PhysRevLett.42.417
  %%CITATION = doi:10.1103/PhysRevLett.42.417;%%
  %395 citations counted in INSPIRE as of 19 Aug 2017
%\cite{Smolin:1979uz}
%\bibitem{Smolin:1979uz}
  L.~Smolin,
  %``Towards a Theory of Space-Time Structure at Very Short Distances,''
  Nucl.\ Phys.\ B {\bf 160} (1979) 253.
  doi:10.1016/0550-3213(79)90059-2
  %%CITATION = doi:10.1016/0550-3213(79)90059-2;%%
  %228 citations counted in INSPIRE as of 19 Aug 2017



\bibitem{unitarity}
%\cite{Giudice:2010ka}
%\bibitem{Giudice:2010ka}
  G.~F.~Giudice and H.~M.~Lee,
  %``Unitarizing Higgs Inflation,''
  Phys.\ Lett.\ B {\bf 694} (2011) 294
  doi:10.1016/j.physletb.2010.10.035
  [arXiv:1010.1417 [hep-ph]].
  %%CITATION = doi:10.1016/j.physletb.2010.10.035;%%
  %92 citations counted in INSPIRE as of 19 Aug 2017


\bibitem{kinterm}
%\cite{Lee:2014spa}
%\bibitem{Lee:2014spa}
  H.~M.~Lee,
  %``Chaotic inflation and unitarity problem,''
  Eur.\ Phys.\ J.\ C {\bf 74} (2014) no.8,  3022
  doi:10.1140/epjc/s10052-014-3022-0
  [arXiv:1403.5602 [hep-ph]].
  %%CITATION = doi:10.1140/epjc/s10052-014-3022-0;%%
  %18 citations counted in INSPIRE as of 19 Aug 2017



\bibitem{linde}
%\cite{Kallosh:2013tua}
%\bibitem{Kallosh:2013tua}
  R.~Kallosh, A.~Linde and D.~Roest,
  %``Universal Attractor for Inflation at Strong Coupling,''
  Phys.\ Rev.\ Lett.\  {\bf 112} (2014) no.1,  011303
  doi:10.1103/PhysRevLett.112.011303
  [arXiv:1310.3950 [hep-th]].
  %%CITATION = doi:10.1103/PhysRevLett.112.011303;%%
  %145 citations counted in INSPIRE as of 19 Aug 2017





\bibitem{riotto}
%\cite{Kehagias:2013mya}
%\bibitem{Kehagias:2013mya}
  A.~Kehagias, A.~M.~Dizgah and A.~Riotto,
  %``Remarks on the Starobinsky model of inflation and its descendants,''
  Phys.\ Rev.\ D {\bf 89} (2014) no.4,  043527
  doi:10.1103/PhysRevD.89.043527
  [arXiv:1312.1155 [hep-th]].
  %%CITATION = doi:10.1103/PhysRevD.89.043527;%%
  %79 citations counted in INSPIRE as of 19 Aug 2017


\bibitem{espinosa}
%\cite{Barbon:2015fla}
%\bibitem{Barbon:2015fla}
  J.~L.~F.~Barbon, J.~A.~Casas, J.~Elias-Miro and J.~R.~Espinosa,
  %``Higgs Inflation as a Mirage,''
  JHEP {\bf 1509} (2015) 027
  doi:10.1007/JHEP09(2015)027
  [arXiv:1501.02231 [hep-ph]].
  %%CITATION = doi:10.1007/JHEP09(2015)027;%%
  %14 citations counted in INSPIRE as of 19 Aug 2017



\bibitem{VSB1}
%\cite{EliasMiro:2012ay}
%\bibitem{EliasMiro:2012ay}
  J.~Elias-Miro, J.~R.~Espinosa, G.~F.~Giudice, H.~M.~Lee and A.~Strumia,
  %``Stabilization of the Electroweak Vacuum by a Scalar Threshold Effect,''
  JHEP {\bf 1206} (2012) 031
  doi:10.1007/JHEP06(2012)031
  [arXiv:1203.0237 [hep-ph]].
  %%CITATION = doi:10.1007/JHEP06(2012)031;%%
  %204 citations counted in INSPIRE as of 19 Aug 2017


\bibitem{VSB2}
%\cite{Lebedev:2012zw}
%\bibitem{Lebedev:2012zw}
  O.~Lebedev,
  %``On Stability of the Electroweak Vacuum and the Higgs Portal,''
  Eur.\ Phys.\ J.\ C {\bf 72} (2012) 2058
  doi:10.1140/epjc/s10052-012-2058-2
  [arXiv:1203.0156 [hep-ph]].
  %%CITATION = doi:10.1140/epjc/s10052-012-2058-2;%%
  %136 citations counted in INSPIRE as of 19 Aug 2017



\bibitem{reheat}
%\cite{Lee:2013nv}
%\bibitem{Lee:2013nv}
  H.~M.~Lee,
  %``Running inflation with unitary Higgs,''
  Phys.\ Lett.\ B {\bf 722} (2013) 198
  doi:10.1016/j.physletb.2013.04.024
  [arXiv:1301.1787 [hep-ph]].
  %%CITATION = doi:10.1016/j.physletb.2013.04.024;%%
  %22 citations counted in INSPIRE as of 19 Aug 2017




\end{thebibliography}
\end{document}